\documentclass[prl, aps, superscriptaddress, amsfonts, twocolumn, showpacs]{revtex4}
\usepackage{graphicx,subfigure,float}
\usepackage{amsmath,amsfonts,bm,graphicx,amssymb,color}
\usepackage[hypertex]{hyperref}

\renewcommand{\d}{\partial}

\newcommand{\lb}{\bigg[}
\newcommand{\rb}{\bigg]}
\newcommand{\lc}{\bigg\{}
\newcommand{\rc}{\bigg\}}

\newcommand{\hatrho}{\hat{\rho}}

\newcommand{\mean}[1]{\langle#1\rangle}
\newcommand{\tr}{\text{Tr }}

\newcommand{\be}{\begin{equation}}
\newcommand{\ee}{\end{equation}}

\begin{document}

\title{Entanglement from Charge Statistics: Exact Relations for Many-Body Systems}
\author{H.~Francis~Song}
\affiliation{Department of Physics, Yale University, New Haven, CT 06520}
\author{Christian~Flindt}
\affiliation{D\'{e}partement de Physique Th\'{e}orique, Universit\'{e} de Gen\`{e}ve, CH-1211 Gen\`{e}ve, Switzerland}
\author{Stephan~Rachel}
\affiliation{Department of Physics, Yale University, New Haven, CT 06520}
\author{Israel~Klich}
\affiliation{Department of Physics, University of Virginia, Charlottesville, VA 22904}
\author{Karyn~Le~Hur}
\affiliation{Department of Physics, Yale University, New Haven, CT 06520}

\begin{abstract}
We present exact formulas for the entanglement and R\'{e}nyi entropies generated at a quantum point contact (QPC) in terms of the statistics of charge fluctuations, which we illustrate with examples from both equilibrium and non-equilibrium transport. The formulas are also applicable to groundstate entanglement in systems described by non-interacting fermions in any dimension, which in one dimension includes the critical spin-1/2 XX and Ising models where conformal field theory predictions for the entanglement and R\'{e}nyi entropies are reproduced from the full counting statistics. These results may play a crucial role in the experimental detection of many-body entanglement in mesoscopic structures and cold atoms in optical lattices.
\end{abstract}

\date{\today}
\maketitle

\begin{figure}
\begin{center}
\includegraphics*[width=215pt]{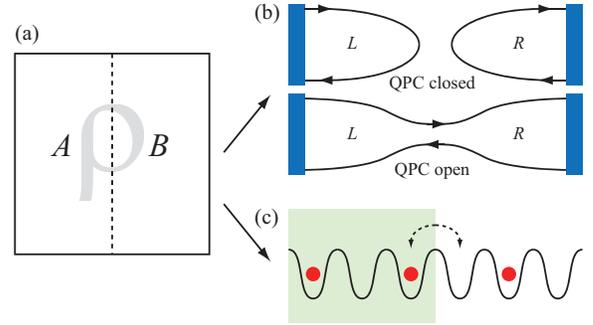}
\caption{\label{fig:setup}(color online). (a) The entanglement entropy characterizes non-local quantum correlations between subsystem $A$ and the remainder $B$, such as (b) two electrodes upon opening of a quantum point contact (QPC) and (c) strongly-repulsive bosons hopping in a one-dimensional optical lattice.}
\end{center}
\end{figure}

\emph{Introduction.}--- Entanglement entropy is playing an increasingly important role in describing quantum correlations in many-body systems \cite{amico-fazio-osterloh-vedral+eisert-cramer-plenio}. For a bipartite system, Fig.~\ref{fig:setup}a, the entanglement entropy of subsystem $A$ is defined as $\mathcal{S}_A =-\tr_{\!\!\! A}\{\hatrho_A \log \hatrho_A\}$, where the reduced density matrix $\hatrho_A=\tr_{\!\!\! B}\{\hatrho\}$ is obtained from the full density matrix $\hatrho$ by tracing out the degrees of freedom in the remainder $B$. For a pure state $\hatrho=|\Psi\rangle\!\langle\Psi|$, $\mathcal{S}_A=\mathcal{S}_B\equiv\mathcal{S}$. Entanglement entropy is currently being studied theoretically in a wide range of systems including quantum critical systems in one \cite{calabrese-cardy,vidal-latorre-rico-kitaev,refael-moore} and higher \cite{metlitski-fuertes-sachdev+kallin-gonzalez-hastings-melko} dimensions, topologically-ordered states \cite{kitaev-preskill+levin-wen}, and evolution after a quench \cite{calabrese-cardy-time+eisert-osborne}. Experimental progress, however, has so far been hindered by the difficulty of measuring the density matrix of a quantum many-body system and the fact that the definition of entanglement entropy itself does not refer to any directly measurable observables.

There has thus been a growing interest in relating entanglement entropy to experimentally accessible quantities, in particular fluctuations of charge and magnetization \cite{klich-refael-silva}. An important step towards this goal was taken by Klich and Levitov who suggested an intimate connection between entanglement entropy and current fluctuations for non-interacting fermions \cite{klich-levitov}. They studied the entanglement entropy between two electronic leads connected via a quantum point contact (QPC), Fig.~\ref{fig:setup}b, and found that it can be expressed as a series in the cumulants of the current fluctuations. Remarkably, for purely gaussian fluctuations with variance $C_2$ their result $\mathcal{S}=(\pi^2/3)C_2=(1/3)\log (\mathcal{T}/\tau)$ reproduces the conformal field theory (CFT) prediction $\mathcal{S}=(c/3)\log (\mathcal{L}/\xi)$ with the spatial extent $\mathcal{L}$ replaced by the temporal window $\mathcal{T}$ during which the QPC is open, the spatial cut-off $\xi$ replaced by the short-time cut-off $\tau$, and the central charge equal to unity $c=1$ \cite{calabrese-cardy,hsu-grosfeld-fradkin,song-rachel-lehur}. 

However, the series developed in Ref.\ \onlinecite{klich-levitov} does not converge in general for non-gaussian fluctuations \cite{flaw}. This constitutes a major obstacle to further systematic study of entanglement and its relation to fluctuations in many-body systems. To illustrate the acuteness of this problem, consider the simplest situation involving a QPC where only a single charge is transmitted with probability $1/2$: The series in Ref.~\onlinecite{klich-levitov} expresses the entanglement entropy as an infinite sum over charge cumulants $C_{n}$, where the coefficients for high-order cumulants asymptotically go to a constant ($\sim 2C_n$ for large orders $n$). A quick calculation shows that for this simple process the $C_{n}$ diverge as $|C_n|\sim 2(n-1)!/\pi^n$, and indeed such factorial divergences are quite typical for non-gaussian distributions \cite{flindt-etal}. Experimental determination of the entanglement entropy from the cumulants using the results of Ref.~\onlinecite{klich-levitov} therefore requires further resummation.

 In this Letter we present exact and convergent series for the entanglement and R\'{e}nyi entropies. The formulas are applicable to a large class of many-body systems, which we illustrate with examples of entanglement generation in a QPC, Fig.\ \ref{fig:setup}b, and groundstate entanglement of systems described by non-interacting fermions in one and two dimensions. In one dimension the latter includes the equivalent, and experimentally relevant, case of strongly-repulsive bosons in an optical lattice, Fig.\ \ref{fig:setup}c, as well as the spin-1/2 XX and Ising models. Our formulas pave the way to an improved understanding of fluctuations and entanglement in quantum many-body systems and may facilitate experimental investigations of entanglement entropy in mesoscopic structures where the measurement of high-order current correlators is becoming possible \cite{reulet-senzier-prober+bomze-gershon-shovkon-levitov-reznikov}.

\begin{figure}[t!]
\hspace{-7pt}
\includegraphics*[width=240pt]{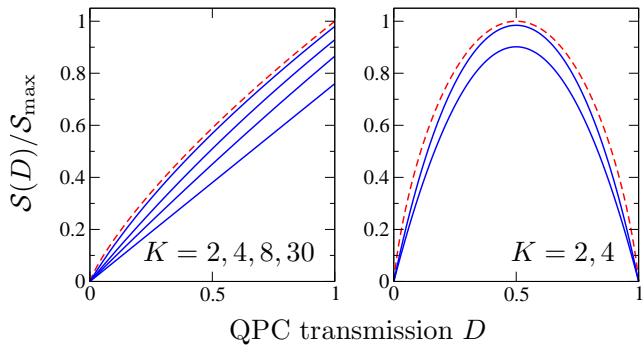}
\caption{\label{fig:bias}(color online). Exact entanglement entropy (dashed) and approximation by cumulants (solid) for a QPC with transmission D for (a) zero bias voltage and (b) bias voltage $V$, scaled to the maximum value at $D=1$ and $D=0.5$, respectively. The cutoff $K$ for the number of cumulants included, from bottom to top, is indicated in each plot.}
\end{figure}

\begin{figure}[t!]
\includegraphics*[width=240pt]{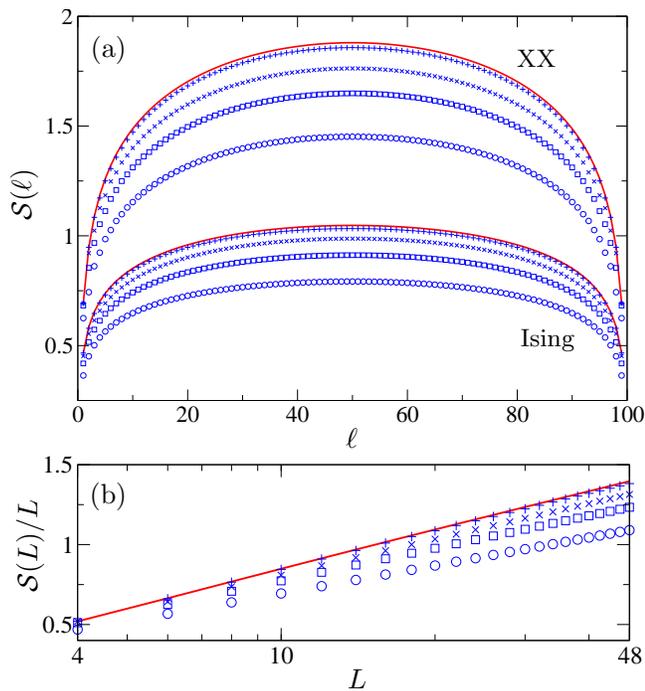}
\caption{\label{fig:ff}(color online). The exact entanglement entropy and approximation by cumulants for (a) the critical spin-1/2 XX chain (upper set of curves) and critical Ising chain in transverse field (lower set of curves) with periodic boundary conditions, $L=100$, and (b) free fermions on $L\times L$ square lattices with periodic boundary conditions, with the partition taken to be half the lattice. Solid curves are the exact entanglement entropy, while the cutoff number $K$ for both figures is 2 (circles), 4 (squares), 8 (crosses), and 40 (plusses).}
\end{figure}

\emph{Main result.}--- We begin by considering the QPC depicted in Fig.\ \ref{fig:setup}b, but we will see below that the result applies to a much larger class of problems. Charge fluctuations in one of the leads of the QPC (the right lead, for concreteness) are characterized by the probability $P_n$ of transmitting $n$ charges from the left to the right lead. The main result of this work relates the entanglement entropy to the charge statistics as
\be
	\mathcal{S} = \lim_{K\rightarrow\infty} \sum_{n=1}^{K+1} \alpha_n(K)C_n, \label{eq:newseries}
\ee where $C_n=(-i\d_\lambda)^n \log \chi(\lambda)|_{\lambda=0}$ are the cumulants of $P_n$ and $\chi(\lambda)= \sum_n P_n e^{i\lambda n}$ is the generating function. The cutoff-dependent coefficients $\alpha_n(K)$ are
\be
	\alpha_n(K) =
	\begin{cases}
	2\sum_{k=n-1}^K \frac{S_1(k,n-1)}{k!k} & \text{for $n$ even},\\
	0 & \text{for $n$ odd},
	\end{cases}\label{eq:newseriescoeff}
\ee
where $S_1(n,m)$ are the unsigned Stirling numbers of the first kind. A beautiful property of the series (\ref{eq:newseries}) is that only even-order cumulants contribute, reflecting the requirement that the entanglement entropy be symmetric between the left and right leads in the particular case of a pure state: Since charge conservation implies that the number of electrons $n$ collected in the right lead is equal to $-n$ charges collected in the left, only even-order cumulants are symmetric in the two leads. Moreover, $\alpha_n(\infty)=2\zeta(n)$ for even $n$ where $\zeta(n)$ is the Riemann zeta function, and in particular we reproduce the CFT result $\mathcal{S}=(\pi^2/3)C_2$ for purely gaussian fluctuations where all cumulants except the first and second are zero. In general, however, the number of cumulants included fixes the cutoff $K$. As $K$ grows the series becomes an increasingly sharper estimate of $S$ from below. %This will be seen to be responsible for the very different convergence properties of Eq.~\eqref{eq:newseries} for non-gaussian fluctuations from the series presented in Ref.\ \onlinecite{klich-levitov}.
This feature of the series is important from a practical point of view, since it ensures that including more terms always improves the estimate, e.g. Figs.~\ref{fig:bias} and \ref{fig:ff}.

\emph{Derivation.}--- To derive Eq.~\eqref{eq:newseries} we begin with the expression for the entanglement entropy
\be
	\mathcal{S} = -\tr \{ M \log M + (1-M) \log (1-M)\},
\label{eq:Sdef}
\ee
where $M=P_R n_U P_R$ is the correlation matrix projected onto the modes in the right lead by $P_R$, and we assume that $M$ is either finite or can be regularized such that the operations below are well-defined \cite{muzykantskii-adamov+avron-bachmann-graf-klich}. Here $n$ is the Fermi-Dirac distribution before the evolution, and $n_U=UnU^\dag$ represents the state of the total system after evolution $U$ of the single particle modes. We first expand the logarithms in Eq.~\eqref{eq:Sdef} around $M=0,1$ to obtain $\mathcal{S}=\sum_{n=1}^\infty A_n/n$ with
\be
	A_n = \tr \{M(1-M)^n + M^n(1-M)\}.
\label{eq:Antrace}
\ee To relate the coefficients $A_n$ to measurable quantities, we next use the Levitov--Lesovik determinant formula \cite{levitov-lesovik} for the generating function of the charge transport statistics written as
\be
\chi(\lambda) = \det [(1-M+Me^{i\lambda})e^{-i\lambda UnP_RU^\dag}].
\label{eq:chi}
\ee 
For such a generalized binomial distribution it is useful to consider the \emph{factorial} cumulants
%noting that the \emph{factorial} cumulants 
$F_n=\d_\lambda^n\log\chi(-i\log\lambda)|_{\lambda=1}$, since for $n\geq 1$ they are related in a simple way to $\tr\{M^n\}$ as $F_n = (-1)^{n-1}(n-1)!(\tr \{M^n\} - \tr\{UnP_RU^\dag\})$. Solving for $\tr\{M^n\}$ and substituting into Eq.~\eqref{eq:Antrace} then gives
\be
	\mathcal{S} = \sum_{n=1}^\infty \lc\frac{(-1)^{n-1}}{n}\lb \frac{F_n}{(n-1)!} + \frac{F_{n+1}}{n!} \rb + \sum_{k=0}^n\binom{n}{k}\frac{F_{k+1}}{k!n}\rc.
\label{eq:factorialseries}
\ee
To write the sum in terms of ordinary cumulants, a cutoff $K$ is introduced so that $\mathcal{S}=\lim_{K\rightarrow\infty}\sum_{n=1}^K A_n/n$. Using the relation $F_n=\sum_{k=0}^n (-1)^{n-k}S_1(n,k)C_k$ between factorial and ordinary cumulants, we finally arrive at Eqs.\ (\ref{eq:newseries}) and \eqref{eq:newseriescoeff} after some algebra.

\emph{Convergence.}--- The convergence of the series \eqref{eq:newseries} for any $M$ is most conveniently shown with the equivalent form (\ref{eq:factorialseries}). The counting statistics for non-interacting electron transport through a two-terminal conductor is always generalized binomial \cite{abanov-ivanov-factorization}, such that the generating function can be factored into a product of binomial events with individual probabilities $0\leq p_i\leq 1$. This is equivalent to evaluating Eq.~\eqref{eq:Antrace} in the eigenbasis of $M$. Then $\mathcal{S}=\sum_i H_2(p_i)$, where $H_2(x)=-x\log x-(1-x)\log (1-x)$ is the binary entropy function. Since $M$ is assumed to be finite and $H_2(x)$ has the convergent (absolutely, by the ratio test) series expansion $\sum_{n=1}^\infty A_n/n$ with $A_n=\sum_i[p_i(1-p_i)^n+p_i^n(1-p_i)]$, the series \eqref{eq:factorialseries} also converges to the correct value. Moreover, since each term $A_n/n$ is positive and decreasing for increasing $n$, Eq.~\eqref{eq:factorialseries} yields an increasingly sharper lower bound to the exact entanglement entropy. The same conclusion holds for the series expressed in terms of ordinary cumulants, i.e., as the cutoff $K$ is increased the sum (\ref{eq:newseries}) {\it converges from below} to the exact entanglement entropy. 

The rate of convergence generally depends on the distribution of eigenvalues $p_i$; specifically, the expansion of the logarithms in Eq.\ (\ref{eq:Sdef}) implies that the presence of eigenvalues of $M$ near 0 or 1 will slow the rate of convergence. As shown below, typically the first few terms dominate the series, but full convergence may require the inclusion of many terms.

\emph{R\'{e}nyi entropies}.--- Before turning to applications of Eq.~\eqref{eq:newseries}, we briefly show that the same approach can also be used to compute the R\'{e}nyi entropies $\mathcal{S}_n\equiv (1-n)^{-1}\log[\tr\{\hatrho_R^n\}]$ of order $n>1$, which can be written as $\mathcal{S}_n = (1-n)^{-1}\tr\{\log[M^n + (1-M)^n]\}$. From a similar derivation to Eq.~\eqref{eq:newseries} we obtain $\mathcal{S}_n = \lim_{R\rightarrow\infty}\sum_{k=1}^{nR}\beta_k(n,R)C_k$, where $R$ is another cutoff and
\be
	\beta_k(n,R) = \begin{cases}
		\frac{1}{1-n}\sum_{r=1}^R\sum_{m=0}^r\sum_{s=k}^{nr} (-1)^{r+s+nr+nm}\\
	\qquad\times\frac{1}{r}\binom{R}{r}\binom{r}{m}\binom{nm}{nr-s}\frac{S_1(s,k)}{(s-1)!} \text{\ \ \ \ for $k$ even},\\
		0 \qquad\qquad\qquad\qquad\qquad\qquad \text{\ \ \ for $k$ odd}.
	\end{cases}\label{eq:renyi}
\ee
Once again only even cumulants contribute to the sum. Moreover, for the gaussian case where all but the first and second cumulants vanish it can be shown that $\beta_2(n,\infty)=(\pi^2/6)(1+1/n)$, which correctly reproduces the CFT result for the R\'{e}nyi entropies \cite{calabrese-cardy-2}.

While we have expressed the entanglement and R\'{e}nyi entropies in terms of the cumulants of the current with experimental applications in mind, it is also useful to express the entropies in terms of the cumulant generating function itself. This can be done by relating both the R\'{e}nyi entropies and the generating function to the spectral density $\mu(z)=\tr \delta(z-M)$ of $M$ to obtain
\be
	\mathcal{S}_n = \frac{n}{\pi}\int_{-\infty}^\infty du\ \frac{\tanh u-\tanh (nu)}{1-n}
	\text{Im}[\log \chi(\pi-2iu)], \label{eq:renyichi}
\ee where the limit $n\rightarrow1$ gives the entanglement entropy as
\begin{align}
	\mathcal{S} = \mathcal{S}_1 = \frac{1}{\pi}\int_{-\infty}^\infty du\ \frac{u}{\cosh^2{u}}\text{Im}[\log \chi(\pi-2iu)]. \label{eq:entangchi}
\end{align} By numerically integrating these expressions we can benchmark Eq.~\eqref{eq:newseries} against the exact results when only a finite number of cumulants are included, Fig.~\ref{fig:bias}.

\emph{Entanglement production in a QPC.}--- As our first application of Eq.~\eqref{eq:newseries} we consider the generation of entanglement due to current fluctuations in a QPC. For the equilibrium setup at zero bias and transmission $D$, the generating function is given by \cite{levitov-lee-lesovik}
\be
	\chi(\lambda) = e^{-\lambda_*^2G/(2\pi^2)}, \qquad \sin\frac{\lambda_*}{2} = \sqrt{D}\sin\frac{\lambda}{2}, \label{eq:imperfectQPC}
\ee with $G=\log\{[h\beta/(\pi\tau)]\sinh[\pi t/(h\beta)]\}$, where $t$ is the time during which the QPC is open, $\tau$ a short-time cutoff, and $\beta=1/(k_BT)$ the inverse temperature. Here $h$ and $k_B$ are the Planck and Boltzmann constants, respectively. For imperfect transmission $D<1$, the generating function \eqref{eq:imperfectQPC} possesses cumulants of all (even) orders, while for $D=1$ the statistics is gaussian. At short times $\tau\ll t \ll h\beta$ the growth of entanglement entropy is logarithmic, while at long times $t \gg h\beta$ the growth is linear. Thus at any non-zero temperature the entanglement production rate eventually becomes constant. If a DC voltage bias $V$ is applied, the non-equilibrium generating function at zero temperature (the quantum shot noise regime) is found to be the binomial distribution $\log\chi(\lambda)=(eVt/h)\log(1-D+De^{i\lambda})$, for which $\mathcal{S}=-(eVt/h)[D \log D + (1-D)\log(1-D)]$ \cite{klich-levitov}. This is also the electron-hole entanglement predicted for a biased tunnel junction \cite{beenakker}, and indeed in both cases the R\'{e}nyi entropies for small $D$ are $\mathcal{S}_n\simeq [n/(n-1)]N$, where $N=(eVt/h)D$ is the number of transferred particles \cite{nazarov}.

The dependence of the entanglement entropy on the transmission $D$ in the two cases is shown in Fig.~\ref{fig:bias}. We see that the first few cumulants give a good approximation to the exact result obtained from Eq.~(\ref{eq:entangchi}), especially for the case of a bias voltage. This indicates that the detection of entanglement generation in a QPC may be within experimental reach using currently available noise measurement techniques \cite{reulet-senzier-prober+bomze-gershon-shovkon-levitov-reznikov}. 

\emph{Groundstate entanglement in the XY model.}--- We next consider a very different setup from the QPC, namely the bipartite entanglement entropy of the groundstates of systems described by free fermions, such as the spin-$1/2$ XY model in one dimension with Hamiltonian
\be
	\hat{H} = J\sum_i [(1+\gamma)\hat{S}^x_i\hat{S}^x_{i+1} + (1-\gamma)\hat{S}^y_i\hat{S}^y_{i+1}] + b\sum_i\hat{S}^z_i. \label{eq:HXX}
\ee This Hamiltonian is equivalent to a problem of free spinless fermions through the Jordan-Wigner transformation \cite{lieb-schultz-mattis} and for $\gamma=0$ also describes bosons in a one-dimensional lattice with infinite on-site repulsion, Fig.\ \ref{fig:setup}c. Here we are concerned with relating the entanglement entropy to the fluctuations as a function of subsystem size $\ell$. The central quantity is $M=(1+\sqrt{G^TG})/2$, where $G_{ij}=\mean{(\hat{a}^\dag_i-\hat{a}_i)(\hat{a}^\dag_j+\hat{a}_j)}$ with the indices $i,j$ restricted to the first $\ell$ sites of the chain and $\hat{a}_i$ is the Jordan-Wigner fermion annihilation operator on site $i$. The entanglement entropy is then given by Eq.~\eqref{eq:Sdef} \cite{vidal-latorre-rico-kitaev}, while the corresponding generating function is $\chi(\lambda,\ell)=\det(1-M+Me^{i\lambda})$ up to an irrelevant phase. For the XX model at $\gamma=0$ the generating function describes fluctuations of $\hat{S}_z$, i.e., $\chi(\lambda,\ell)=\mean{\exp(i\lambda \sum_{i=1}^\ell \hat{S}_i^z)}=\mean{\exp(i\lambda \sum_{i=1}^\ell (\hat{a}^\dag_i\hat{a}_i-1/2))}$, while for $\gamma\neq0$ the fluctuations must be interpreted as those of quasi-particles, since total $\hat{S}^z$ is not a conserved quantity. The results for both the critical XX model at zero magnetic field ($\gamma=0,b=0$) and critical Ising model ($\gamma=1,b=J$) with periodic boundary conditions are shown in Fig.~\ref{fig:ff}a. Remarkably, the cumulants correctly reproduce the two central charges $c=1$ and $c=1/2$ for the XX and Ising models, respectively \cite{vidal-latorre-rico-kitaev}. We also note that the random singlet phase \cite{refael-moore} can be studied with this method by randomly varying $J$ and $b$ across different sites.

More generally, the same formalism is applicable to any quadratic fermionic Hamiltonian, regardless of dimension, and as such can also be used to determine the entanglement entropy from the fluctuations in higher dimensions. On the $L\times L$ square lattice in two dimensions with nearest-neighbor hopping, for example, we find that the approximation by cumulants follows the predicted $\mathcal{S}\sim L\log L$ scaling \cite{gioev-klich+wolf} as shown in Fig.~\ref{fig:ff}b with the subsystem taken to be half the lattice.
%More generally, the same formalism may be applied to the quadratic Hamiltonian $\hat{H}=\sum_{ij} [\hat{c}^\dag_iA_{ij}\hat{c}_j + (\hat{c}^\dag_iB_{ij}\hat{c}^\dag_j + \text{h.c.})]$ in higher dimensions. Here $\hat{c}_i$ are fermionic annihilation operators while $A$ and $B$ are symmetric and antisymmetric matrices, respectively, with $B\neq0$ corresponding to Hamiltonians of the BCS type. On the $L\times L$ square lattice in two dimensions with $B=0$ and $A$ representing nearest-neighbor hopping, for example, we find the predicted $\mathcal{S}\sim L\log L$ scaling \cite{gioev-klich}, with the approximation by cumulants behaving the same way as shown in Fig.~\ref{fig:ff}b with the subsystem taken to be half the lattice.

\emph{Conclusions.}--- We have derived exact formulas for the entanglement and R\'{e}nyi entropies in terms of the statistics of charge fluctuations. The expressions are applicable to a wide range of experimentally relevant many-body systems, for example entanglement produced in QPCs and groundstate entanglement of systems described by non-interacting fermions (or quasi-particles) in any dimension.
%together with Hamiltonians of the BCS type. 
% It is hoped that the present work will lead to experimental verification of CFT predictions. 

\emph{Acknowledgements.}--- HFS and KLH acknowledge support by NSF Grant No. DMR-0803200 and the Yale Center for Quantum Information Physics (DMR-0653377). The work by CF was supported by the Carlsberg Foundation, and SR acknowledges support from the Deutsche Forschungsgemeinschaft under Grant No. RA 1949/1-1. IK and KLH thank the Aspen Center for Physics for its kind hospitality.

\bibliographystyle{h-physrev}

\begin{thebibliography}{10}
\bibitem{flaw}
Briefly, the problematic step in Ref.~\onlinecite{klich-levitov} involves integrating the cumulant generating function $\log \chi(\lambda)=\sum_{n=1}^\infty [(i\lambda)^n/n!]C_n$ over values of $\lambda$ that extend beyond the radius of convergence, since typical (i.e., non-gaussian) cumulant generating functions have singularities in the complex plane.
\end{thebibliography}

\end{document}